\documentclass[prb,preprint]{revtex4}
\usepackage {graphicx}

\begin{document}

\title{Change in the room temperature magnetic property of ZnO upon Mn doping}

\author{S. Banerjee$^a$\footnote{Email:sangam.banerjee@saha.ac.in},K. Rajendran$^b$,N. Gayathri$^c$\footnote{Present address: Material Science Section, Variable Energy Cyclotron Center, 1/AF Bidhannagar, Kolkata 700 064, India} , M.~Sardar$^c$, S. Senthilkumar$^d$ and V. Sengodan$^e$}

\address{$^a$ Surface Physics Division,  Saha Insitute of Nuclear Physics, 1/AF Bidhannagar, Kolkata 700 064, India\\ $^b$ Department of Electronics and Communication Systems, Sri Krishna Arts and Science College, Coimbatore - 641008, India\\ $^c$ Material Science Division, Indira Gandhi Center for Atomic Research, Kalpakkam 603 102, India\\ $^d$ Department of Chemistry, PSG College of Technology, Coimbatore - 641004, India\\ $^e$ Department of Electronics, SNR Sons College, Coimbatore - 641006, India}

\begin{abstract}
We present in this paper the changes in the room temperature magnetic property of ZnO on Mn doping prepared using solvo-thermal process. The zero field cooled (ZFC) and field cooled (FC) magnetisation of undoped ZnO showed bifurcation and magnetic hysteresis at room temperature. Upon Mn doping the magnetic hysteresis at room temperature and the bifurcation in ZFC-FC magnetization vanishes. The results seem to indicate that undoped ZnO is ferromagnetic while on the other hand the Mn doped ZnO is not a ferromagnetic system. We observe that on addition of Mn atoms the system shows antiferromagnetism with very giant magnetic moments.
  
\end{abstract}

\maketitle

\section{Introduction}
Recently ferromagnetism has been observed in several undoped wide band gap semiconductors such as ZnO, HfO$_2$, SnO$_2$, In$_2$O$_3$, Al$_2$O$_3$, CeO$_2$ Ga$_2$O$_3$ etc. \cite{SBZNO,VenkatesanNature,CoeyPRB,HongPRB2006,Yoon,HongAPL2006,Schwartz2004,Radovanovic,HongPRB2005,CNR,HongJPC2007,Sreedharan}. The observation of ferromagnetism in nonmagnetic metal oxides which are generally supposed to show diamagnetism puzzled us and made us wonder as to what is the origin of formation of moments in these systems. It was conjectured that clustering of oxygen defects in ZnO \cite{SBZNO} and needle like-formation of oxygen O(3) vacancies \cite{Sreedharan} in Ga$_2$O$_3$ are responsible for the formation of the moments. But it is still not very clear why and how these moments collectively forms ferromagnetic ordering.  If some sort of organization of the oxygen vacancies in these system via double exchange or super exchange can give rise to ferromagnetism \cite{condmatznomn} then there is another puzzle: What is then the role of the transition metal (TM) dopant which is generally an important component in making dilute magnetic semiconductor (DMS) from these type of wide band gap oxide systems? Some interesting observation reported earlier are: (1) The total magnetisation observed in Mn doped ZnO reported were sometime as low as $5$x$10^{-3}$ emu/gm \cite{Garcia,Wang,Sharma} but similar values are also observed for  undoped ZnO samples \cite{SBZNO}. (2) If the Mn doped ZnO sample is annealed in air then the net magnetisation of the sample decreases and if annealed in inert atmosphere then the magnetisation of the sample increases\cite{Chen,Hou}. This clearly indicates that annealing the oxides in the presence or absence of oxygen modifies the magnetic property of the material. (3) The controversy still remains whether doping of ZnO with Mn makes the system ferromagnetic. Some reports claim no evidence of ferromagnetism and have rather reasoned it out to be antiferromagnetic near room temperature \cite{Ram,Alaria,Fukumura,Kolesnik}. The nature of magnetic ordering is still to be confirmed in Mn doped ZnO and other similar types of TM doped metal oxide materials. In this paper we have systematically carried out magnetic measurement for  undoped ZnO and as a function of Mn doping, with the main interest to investigate the role of Mn in ZnO in modifying the magnetic property. 

\section{Experimental details}
For the present investigation the  undoped and Mn doped ZnO samples were prepared using solvo-thermal process. The details of the sample preparation and its structural and chemical characterizations are presented elsewhere \cite{condmatchar}. We prepared two sets of samples with each set containing four samples ie., undoped ZnO (with no manganese) and 1 mole\%, 2 mole\% and 3 mole\% of Mn doped in ZnO (henceforth we shall refer to these as 1\%, 2\% and 3\% respectively). These set of samples were annealed at 500$^o$C and 900$^o$C for 1 hour to remove the organic contaminations. The Mn doped samples annealed at 500$^o$C will be called 'Set A' and the Mn doped samples annealed at 900$^o$C will be called 'Set B'. Fig.~1 shows the x-ray patterns of the 3 mole\% (highest doping) Mn doped samples annealed at 500$^o$C and 900$^o$C. The 500$^o$C sample shows no obvious extra peaks, whereas the 900$^o$C sample shows a few extra peaks. The FTIR spectrum of the Mn doped ZnO samples shows (Zn,Mn)-O stretching mode \cite{condmatchar} below 500 cm$^{-1}$. The band-gap obtained from the optical transmittance (UV-VIS-NIR measurements) decreases as the Mn concentration of ZnO increases \cite{condmatchar}. These two results indicate the experimental evidence of Mn doping in ZnO and also indicates that Mn has substituted Zn in the lattice. The magnetic property of the samples were measured using MPMS-7 (Quantum Design). For the zero field cooled (ZFC) data, the sample was cooled down to 2K in the absence of magnetic field and the data was taken while warming up upto 300K in the presence of 100 Oe field. The field cooled data (FC) was taken while warming the sample after cooling it to 2K in the presence of the 100 Oe field. We shall present the hysteresis data taken at 300K with field of upto 2000~Oe for  undoped ZnO sample and upto 7 Tesla for Mn doped ZnO samples. 

\section{Results}
In fig.~2 we show the (M vs. H) hysteresis curves measured at 300K upto 2000~Oe for the  undoped ZnO samples annealed at 500$^o$C and 900$^o$C after the necessary diamagnetic subtraction. We observe clear hysteresis and magnetization saturation at 300K for both the samples.  For the 900$^o$C annealed sample, the saturation magnetisation is higher than that of the 500$^o$C annealed  undoped ZnO sample. The insets of fig.~2 shows the ZFC and FC magnetisation curve taken at 100 Oe for the 500$^o$C and 900$^o$C annealed samples. The distinct bifurcation of the ZFC and FC starts well above the room temperature i.e., 340K similar to the result reported recently in  undoped ZnO sample synthesised using micellar method \cite{SBZNO}. 

In fig.~3(a) we show the ZFC and FC magnetisation curve taken at 100 Oe for the Mn doped ZnO samples (Set A) annealed at 500$^o$C. The distinct bifurcation of the ZFC and FC around room temperature can only be seen for the 1$\%$ Mn doped sample and as the percentage of the Mn increases i.e., 3$\%$ Mn doped, the distinct bifurcation of the ZFC and FC is seen only around ~40K (see inset of fig.~3(a)) and beyond this temperature both the ZFC and FC curve almost overlap each other. For 2$\%$ Mn doped no distinct bifurcation of the ZFC and FC is seen. In fig. 4(a) we show the M vs. H curve at 300K for all the doped samples annealed at 500$^o$C.  From the 300K M vs. H curve (fig. 4(a)) we see almost a linear M vs. H curve except at very low fields where we observe non-linearity. We observe no hysteresis at 300K for all the samples unlike the  undoped ZnO sample annealed at 500$^o$C. The non-linearity behaviour is more for the doping concentration of 1$\%$ Mn than the higher doped samples. This indicates that for the lower concentration of Mn doped sample (i.e.,1$\%$ Mn doped) we still see the remanescent of the magnetic nature of the  undoped ZnO sample but on increasing the doping concentration, this magnetic remanescent gets diminished even though the net magnetisation of the samples increases due to the strong paramagnetic contribution due to Mn doping. Thus for 1$\%$ Mn doping we see bifurcation of ZFC and FC above room temperature due to the presence of remanescent magnetism of undoped ZnO.

In fig.~3(b)  we show the ZFC and FC magnetisation curve taken at 100 Oe for the Mn doped ZnO samples annealed at 900$^o$C (Set B). For this set we see the distinct bifurcation of the ZFC and FC around ~35 K for all the samples (i.e., 1$\%$, 2$\%$ and 3$\%$ Mn in ZnO) and beyond this temperature both the ZFC and FC curve almost falls on each other. A peculiar behaviour i.e., appearance of shoulder around 35K in both the ZFC and FC curves for the 900$^o$C annealed Mn doped ZnO samples is to be noted. This shoulder has earlier been observed in Mn doped ZnO systems \cite{Fukumura,Han2,Chen2}. We also performed M vs. H measurement at 300K for all the doped samples and this is shown in the fig~4(b). The 300K  curves show no hysteresis. From the M vs. H curve at 300K (fig. 4(b)) we see perfect linear M vs. H curve even down to very low fields (see the inset). The linear M vs. H curves indicate that the samples appeares to show perfect paramagnetism around room temperature. This reveals that on doping ZnO with Mn the room temperature magnetic hysteresis of the ZnO gets quenched completely as a function of Mn concentration. In the next section we shall try to understand the exact magnetic nature of the Mn doped samples using a phenomenological approach.

\section{Analysis and Discussion}
For the analysis of our data we have plotted in fig.~5 the inverse magnetic susceptibility (1/$\chi$) as a function of temperature for T $>$ 50K (i.e., well above the bifurcation temperature) obtained from the magnetization vs temperature data for both the set of samples (ie., 500$^o$C and the 900$^o$C annealed) except the 1 mole\% Mn 500$^o$C annealed sample because the sample still shows the ferromagnetic remnescence of the pure ZnO samples as mentioned earlier. At high temperatures we observe that the inverse of susceptibility depends linearly on temperature with a negative intercept  for all the samples shown in fig.~5. The negative intercept in the temperture is a signature of antiferromagnetic coupling in the samples.  At lower temperatures the inverse susceptibity deviates from its linear nature and approaches zero as the temperature (T) tends to zero which can happen only if there is a paramagnetic moments also in the samples. Using a phenomenological approach we can express the susceptibility as a sum of a Curie (paramagnetic-like) and Curie-Weiss (antiferromagnetic-like) terms i.e.,

\begin{equation}
\chi=C[{f\over T}+{(1-f)\over T+\theta }]
\end{equation}
where,
\begin{equation}
C={N g^2\mu_B^2 S(S+1) \over 3k_B}
\end{equation}

\noindent where, $N$ is number of Mn atoms/gram,  $g=2.0$, $\mu_B=9.27 \times 10^{-21}$ ergs/Oerstead, and $k_B=1.38 \times 10^{-16}$ ergs/Kelvin and $f$ is fraction of Mn atoms that do not have any near neighbour Mn atom, i.e they are paramagnetic. We have fitted the inverse of susceptibility using the above expression as a function of temperature with $f$, $C$ and $\theta$ as the fit parameters. From the fit parameter $C$ we could extract $S(S+1)$  as shown in table I.

The nearest neighbour Mn-Mn exchange constant J is related to Curie-Weiss temperature $\theta$ by \cite{Spalek}

\begin{equation}
{J\over k_B }= {3\over 2} {\theta \over z S(S+1)x}
\end{equation}

\noindent where, $zx$ is the average number of nearest neighbour of Mn atoms around any given Mn site for $x$ mole fraction of the Mn and $z$ = 12 is the number of nearest neighbour for the wurtzite lattice. The obtained value of $J\over k_B$, $\theta$ and $S$ are shown in table I. 

For the 500$^o$C annealed samples (Set A) we observe a lower paramagnetic fraction than the 900$^o$C annealed samples (Set B). For the set A samples the paramagnetic fraction $f$ does not change much while going from $x$ = 0.02 to 0.03. The spin value and the Curie-Weiss temperature are much larger for samples of Set A than that of Set B. For samples of Set B we see a decrease in the paramagnetic fraction with increase in $x$ and the Curie constant $C$ does not scale proportionally with $x$. The $\theta$ value increases with $x$ but not linearly proportional to $x$ as proposed by Spalek et. al. \cite{Spalek}. The spin values of the elementary moments varies from 6.67 to 11.8. These are very large compared to $S=5/2$ for Mn$^{2+}$ ions. Using the large spin values extracted from the Curie constant, the near neighbour average exchange coupling J/$k_B$ between the Mn ions is between 30K to 100K (see Table I). Thus we observe that in our method of sample preparation: (1) There always exist a small paramagnetic fraction of the Mn spins (varying from 5\% to 20\% of the total Mn spins). This small paramagnetic fraction is responsible for the deviation from linearity of 1/$\chi$ at lower temperatures. (2) The dominant interaction between the Mn spins is antiferromagnetic as obtained from the $\theta$ value. (3) The spin values $S$ extracted from the experimental data is always much larger than the expected theoretical values. The spin values seem to depend on the annealing conditions, for example samples annealed at lower temperature (Set A) shows larger S and $\theta$ values than samples of Set B. (4) Even though the Curie-Weiss temperature obtained for these samples are very large we do not observe any spin ordering at any finite temperature. These large values of $\theta$ were also noted by many previous investigators \cite{Ram,Luo}. The large $\theta$ values indicate that either the J value is large or the spin moment $S$ is large. Since we do not observe any spin ordering at any finite temperature we can conclude that the large $\theta$ value is due to large $S$ only. The most interesting question is why do we obtain such large $S$ values. We would like to point out here that this large value of $S$ has also been observed by others in the DMS systems \cite{Dhar,Orlov,Ogale,Song}. One possible way for  formation of large moments is by clustering of Mn atoms. There are indications from theoretical calculation that transition metal atoms form clusters when doped in high band gap semiconductors \cite{Priya}. Since clustering is an activated process one would expect clustering of Mn atoms in higher temperature annealed samples. In contrary, we get larger spin values (from Curie constant $C$) for the lower temperature annealed sample (set A). 

In the following we shall try to understand the strange ferromagnetism in undoped ZnO and the destruction of this ferromagnetism and the appearance of antiferromagnetism upon Mn doping. We shall also try to understand the strong dependence of the ferromagnetic moments in undoped ZnO and unusually large spin values of the Mn doped ZnO samples on annealing temperatures.

First of all the most puzzling question is why does undoped transition metal oxide shows magnetisation hysteresis which has been attributed by various authors to ferromagnetism. Recently \cite{SBZNO} we have argued that even if a single isolated oxygen neutral vacancy cannot have a moment, a cluster of such vacancies is likely to develop a net moment. When these clusters interact with each other either directly or indirectly through isolated oxygen vacancies having paramagnetic moments one can observe some magnetic ordering which can show its signature as magnetic hysteresis. At a higher annealing temperature larger oxygen vacancy clusters can form by the coalescence of smaller clusters. The increase in saturation magnetization observed for the 900$^o$C can be attributed to formation of larger oxygen vacancy clusters.

From the analysis of the susceptibility data of the doped samples it is clear that eventhough there is no magnetic ordering the system is dominated by antiferromagnetic interactions. The most important observation from our ananlysis of susceptibiltiy data is the appearance of  large spin values (large $C$ and $\theta$). This was observed to be very systematic. We also observe that the spin values decrease with increasing annealing temperature. This can happen if the oxygen vacancies as well as the Mn moments are distributed more homogeneously in samples of set A than in set B. The singly charged oxygen vacancy site electron can couple antiferromagnetically with near neighbours Mn ions. This entity could be like a bound magnetic polaron with a large moment centered around the oxygen vacancy site \cite{Song}. The system thus can be thought of as a collection of large moment polarons. This explains the large spin values deduced from the Curie constant. These magnetic polarons interact antiferromagnetically among themselves. The samples of set B have lesser number of such magnetic polarons because many of the oxygen vacancies can form vacancy clusters upon high temperature annealing and hence the magnetic polarons are more well seperated than in samples of set A.  The samples of Set A will have a higher density of the singly charged oxygen vacancies than samples of set B where the oxygen vacancies form clusters. Because of the high densiity in set A, the magnetic polaron formed will be in closer proximity thus causing a overlap of the polarons, resulting in a bigger magnetic polaron entity with a larger spin moment\cite{Song}. These also interact antiferromagnetically but with a larger moment. Similar large moment was observed in Co doped ZnO and also has been attributed to bound magnetic polarons \cite{Song}. 

\section{Conclusion}
In conclusion, we observed ferromagnetic behaviour of  undoped ZnO sample prepared by solvo-thermal (sol-gel) technique. Upon Mn doping we observe quenching of ferromagnetic hysteresis and the magnetisation varies linearly with field (in most cases) but the inverse susceptibility data does not follow a linear dependence on temperature. To fit the inverse susceptibility data we considered antiferromagnetic interaction as an additive term to the paramagnetic contribution. From the inverse susceptibiltiy data we observe large Curie constant values and also giant magnetic moments in Mn doped ZnO samples. The giant magnetic moments were attributed due to the formation of bound magnetic polaron rather than clustering of Mn atoms as suggested by other theoretical studies in these type of systems. These magnetic polarons interact antiferromagnetically among themselves giving rise to the weak antiferromagnetic contribution to the susceptibility.

\newpage

Table 1: Parameters obtained by fitting the inverse susceptibility using eq. (1)- (3).
\begin{table*}
\begin{tabular}{|c|c|c|c|c|c|c|}
\hline
Sample & f(Paramagnetic fraction)  & Curie Constant C & S(S+1) & $\theta$ & J/$k_B$(degree Kelvin) & S \\
\hline
25 & 0.0562 & 0.0335 & 268.0 & 1825.0 & 43 & 15.8\\
\hline
35 & 0.0539 & 0.0366 & 195.0 & 1827.0  & 39 & 13.47\\
\hline
19 & 0.21 & 0.0032 & 51.2 & 410.0 & 100.0 & 6.67 \\
\hline
29 & 0.088 & 0.0092 & 73.6 & 814.0 & 69 & 8.09 \\
\hline
39 & 0.12 & 0.0288 & 153.0 & 660.0  & 28 & 11.8 \\
\hline
\end{tabular}
\end{table*}

\newpage

{\bf Figure Captions}

{\bf Figure 1} XRD pattern of the 3 mole \% samples annealed at 500$^o$C and 900$^o$C. The phase marked as * could be identified as Zn$_2$MnO$_4$ \cite{JAP97} and \# as Mn$_2$O$_3$ \cite{Garcia} in the 900$^o$C annealed sample. The phase marked as 'x' could not be identified.

{\bf Figure 2} Magnetisation (M) versus magnetic field (H) curves for  undoped ZnO sample annealed at 500$^o$C and 900$^o$C. Inset: ZFC and FC magnetisation curves of the (a) 500$^o$C annealed sample taken at 100 Oe field and (b) 900$^o$C annealed sample taken at 100 Oe field.

{\bf Figure 3} ZFC and FC magnetisation curves of the Mn doped  (a) 500$^o$C and (b) 900$^o$C annealed samples taken at 100 Oe field The insets show the expanded region and the bifurcation between the ZFC and FC is indicated by an arrow.

{\bf Figure 4} Magnetisation (M) versus magnetic field (H) curves for 1$\%$, 2$\%$ and 3$\%$ of Mn doping in ZnO sample annealed at (a) 500$^o$C and (b) 900$^o$C  measured at T = 300K upto a magnetic field of 7 Tesla. Insets show the expanded low field region of the respective M vs H curves.

{\bf Figure 5} Inverse magnetic susceptibility (1/$\chi$) as a function of temperature for T $>$ 50K (i.e., well above the bifurcation temperature) obtained from the magnetization vs temperature data (fig. 4) for both the set of samples (a) 500$^o$C and (b) 900$^o$C annealed 

\newpage

\begin{figure}
\includegraphics*[width=12cm,angle=270]{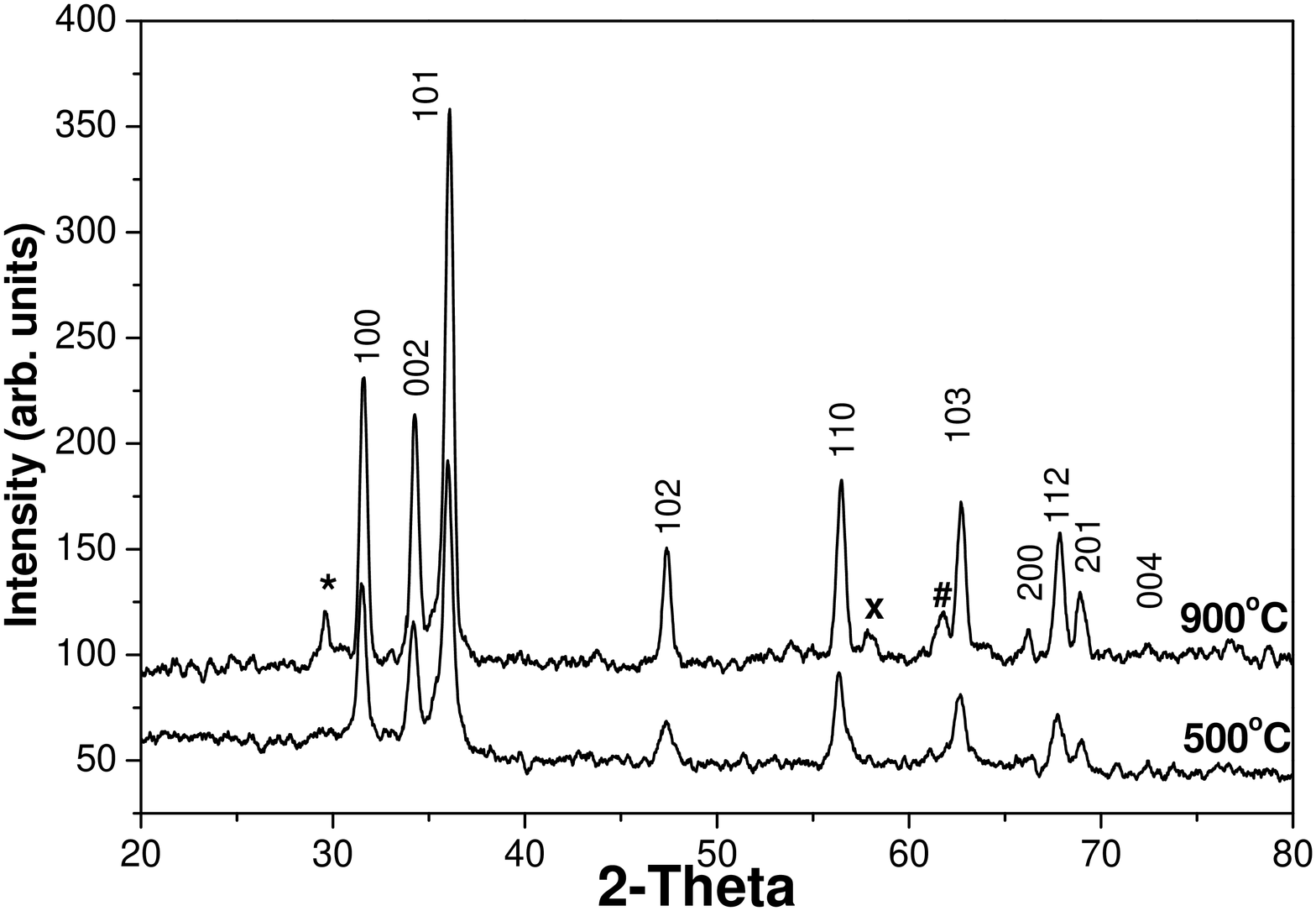}
\caption{XRD pattern of the 3 mole \% samples annealed at 500$^o$C and 900$^o$C. The phase marked as * could be identified as Zn$_2$MnO$_4$ \cite{JAP97} and \# as Mn$_2$O$_3$ \cite{Garcia} in the 900$^o$C annealed sample. The phase marked as 'x' could not be identified.} 
\end{figure}

\begin{figure}
\includegraphics*[width=12cm,angle=270]{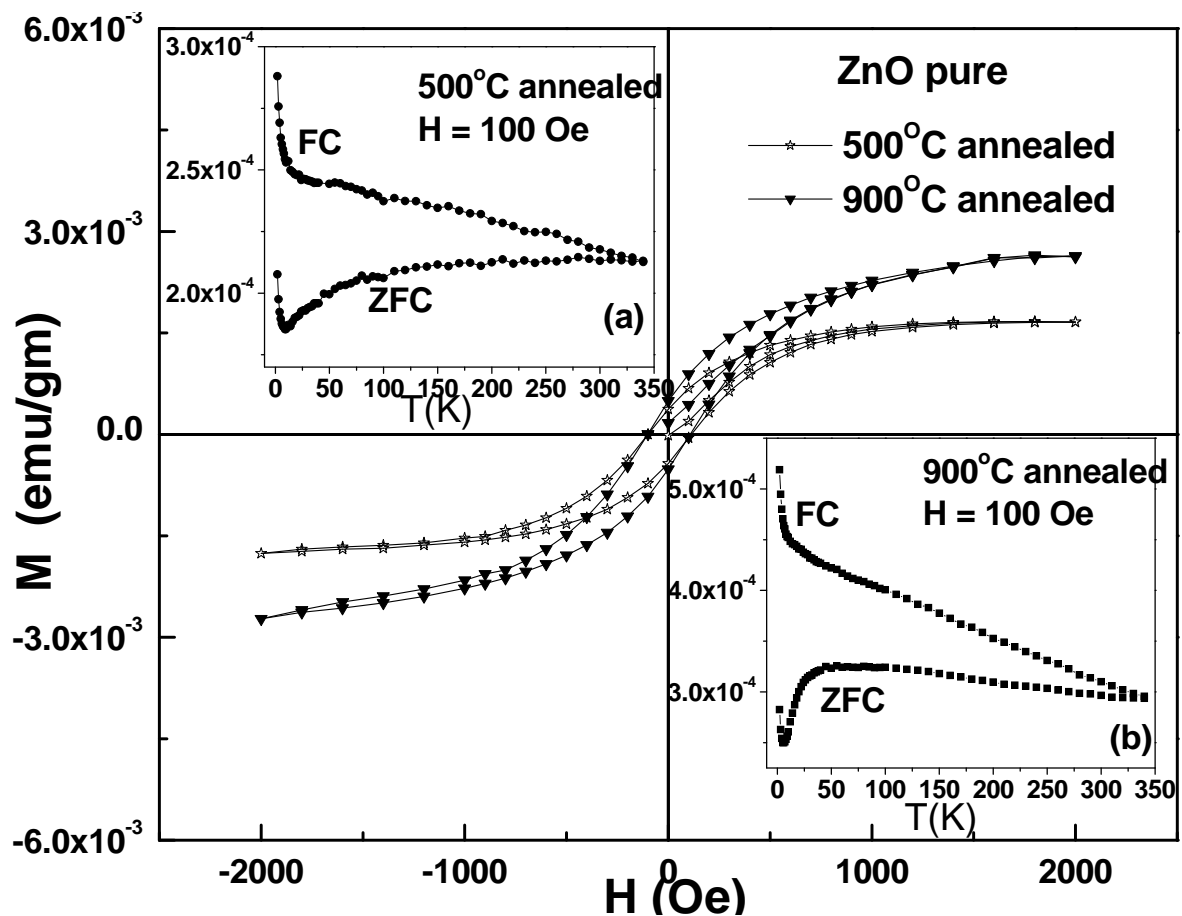}
\caption{Magnetisation (M) versus magnetic field (H) curves for  undoped ZnO sample annealed at 500$^o$C and 900$^o$C. Inset: ZFC and FC magnetisation curves of the (a) 500$^o$C annealed sample taken at 100 Oe field and (b) 900$^o$C annealed sample taken at 100 Oe field}
\end{figure}

\begin{figure}
\includegraphics*[width=12cm]{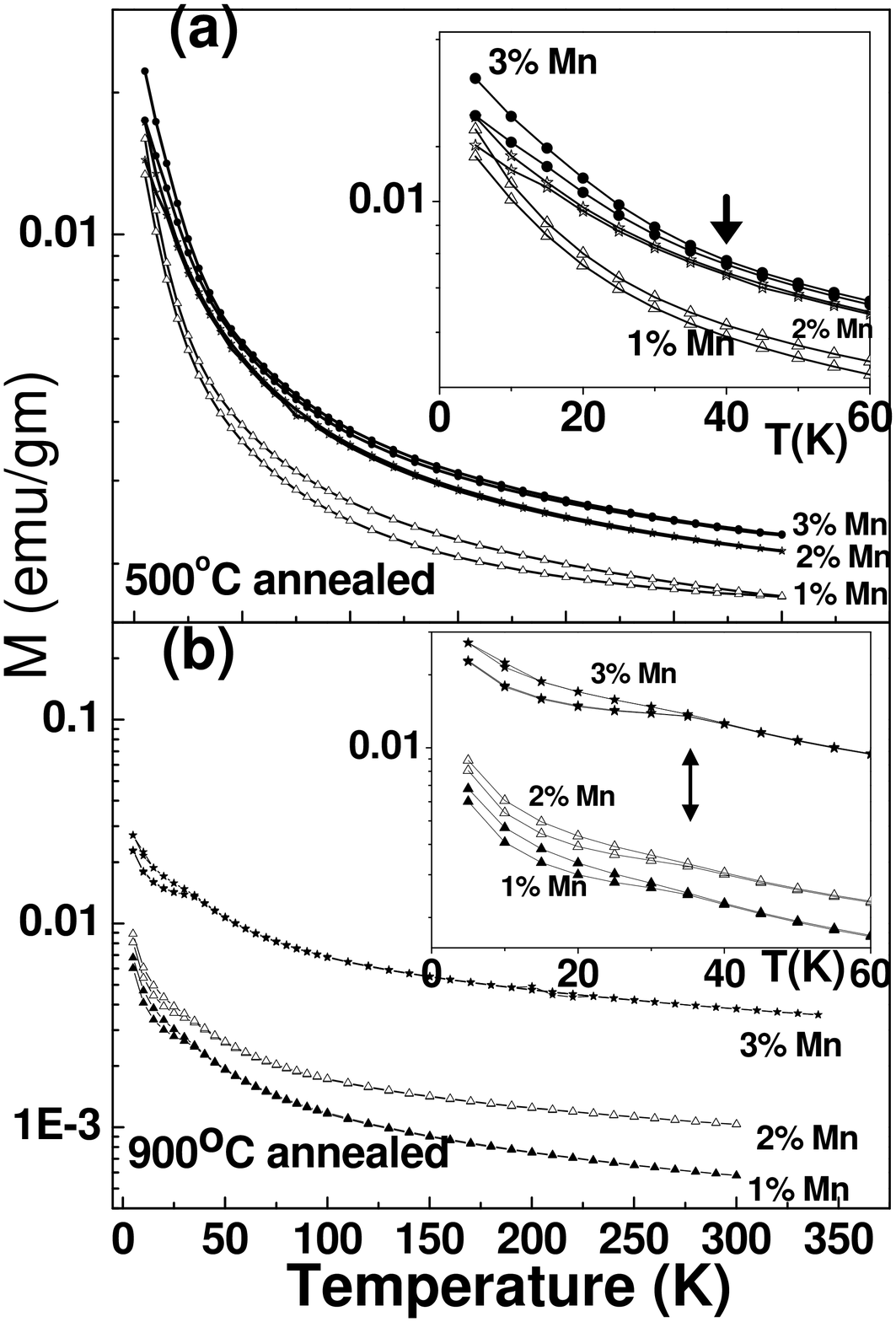}
\caption{ZFC and FC magnetisation curves of the Mn doped  (a) 500$^o$C and (b) 900$^o$C annealed samples taken at 100 Oe field The insets show the expanded region and the bifurcation between the ZFC and FC is indicated by an arrow.}
\end{figure}

\begin{figure}
\includegraphics*[width=12cm]{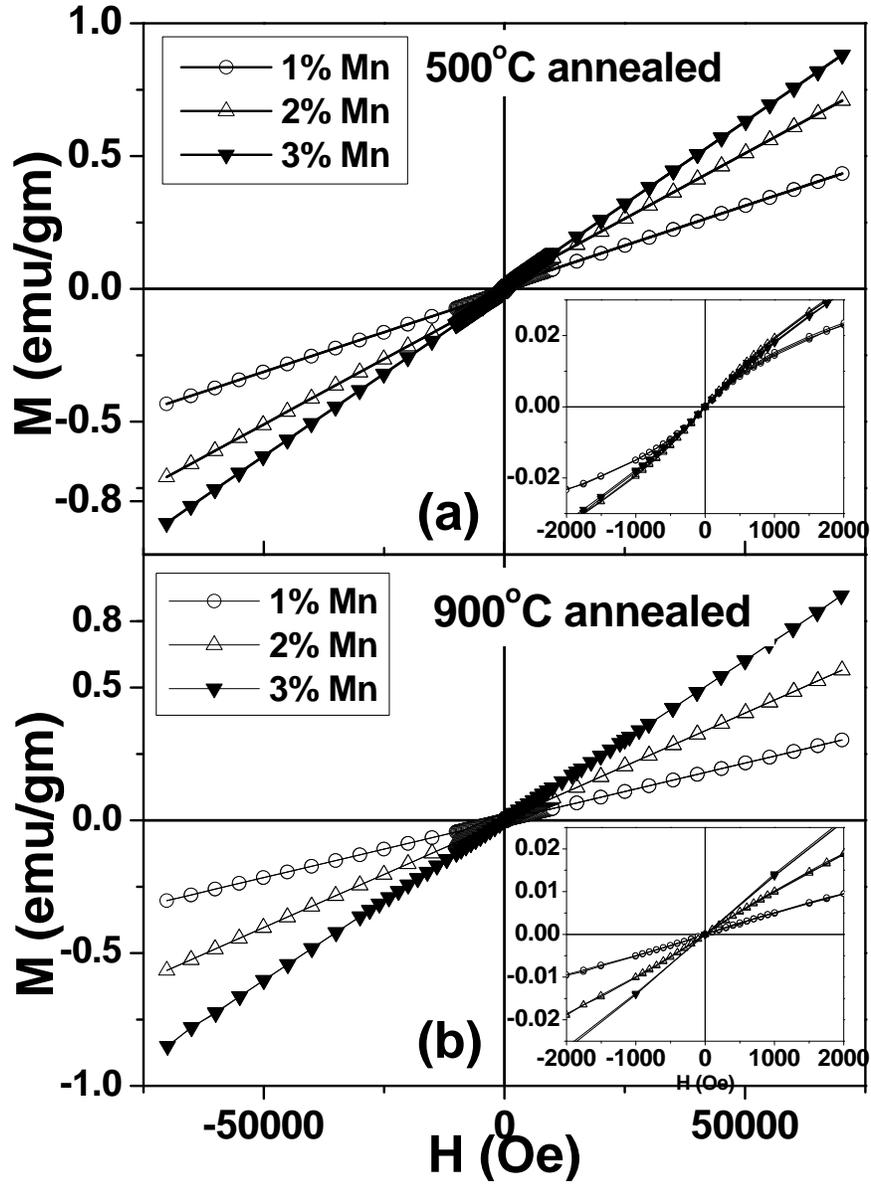}
\caption{Magnetisation (M) versus magnetic field (H) curves for 1$\%$, 2$\%$ and 3$\%$ of Mn doping in ZnO sample annealed at (a) 500$^o$C and (b) 900$^o$C  measured at T = 300K upto a magnetic field of 7 Tesla. Insets show the expanded low field region of the respective M vs H curves.}
\end{figure}

\begin{figure}
\includegraphics*[width=12cm]{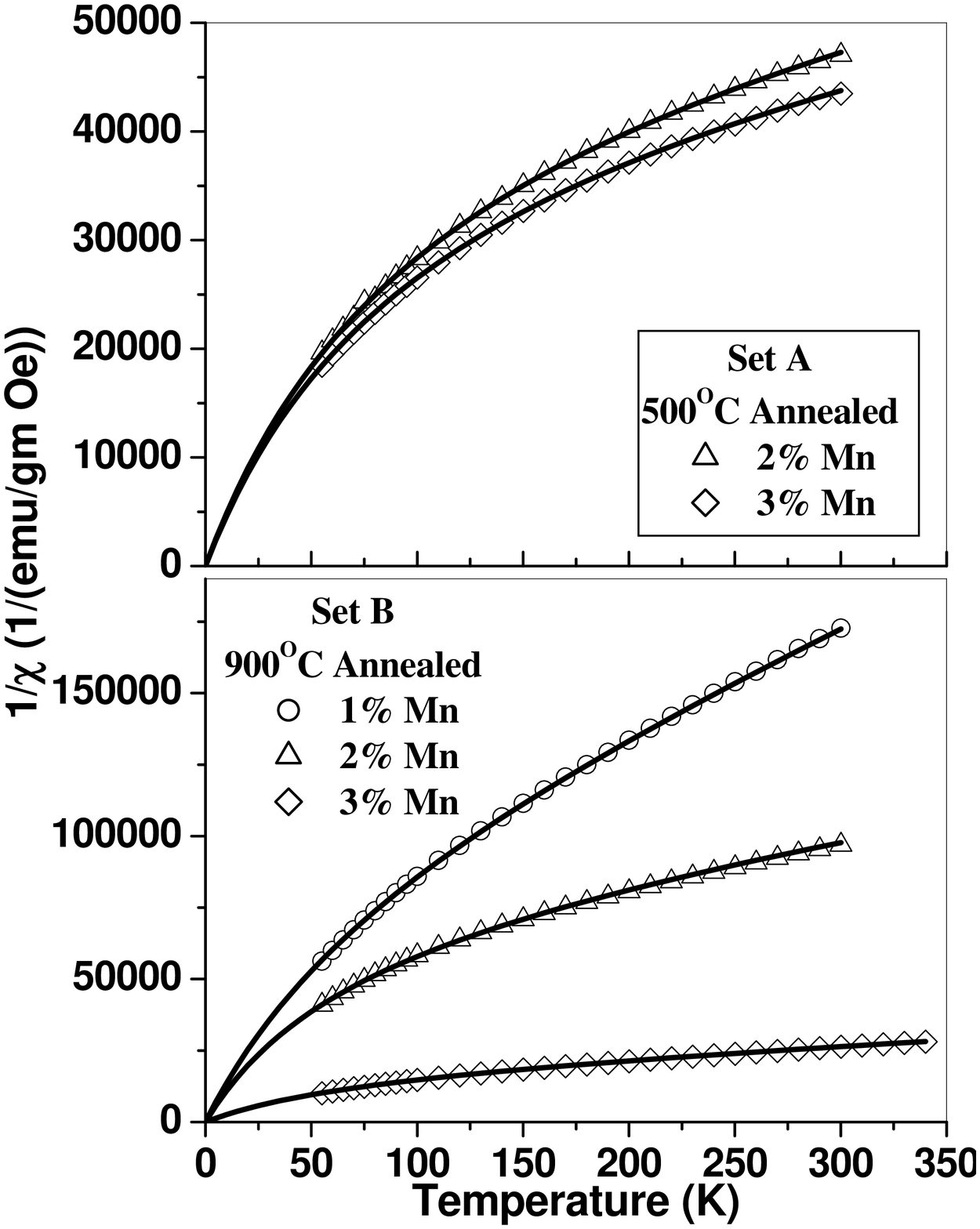}
\caption{Inverse magnetic susceptibility (1/$\chi$) as a function of temperature for T $>$ 50K (i.e., well above the bifurcation temperature) obtained from the magnetization vs temperature data (fig. 4) for both the set of samples (a) 500$^o$C and (b) 900$^o$C annealed }
\end{figure}


\begin{thebibliography}{99}
\bibitem{SBZNO}S. Banerjee, M. Mandal, N. Gayathri and M.Sardar, cond - mat/0702486, cond - mat/0702486v2, Appl. Phys. Letts {\bf91}, 182501 (2007)
\bibitem{VenkatesanNature} M. Venkatesan, C. B. Fitzgerald and J. M. D. Coey, Nature {\bf 430} 630 (2004)
\bibitem{CoeyPRB} J. M. D. Coey, M. Venkatesan , P. Stamenov, C. B. Fitzgerald and L. S. Dorneles, Phys. Rev. B {\bf 72} 24450 (2005)
\bibitem{HongPRB2006}  N. H. Hong, J. Sakai, N. Poirot and V. Brize, Phys. Rev. B {\bf 73} 132404 (2006)
\bibitem{Yoon} S. D. Yoon, Y. Chen, A. Yang, T. L. Goodrich, X. Zou, D. A. Arena, K. Ziemer, C. vittoria and V. G. Harris, J. Phys: Condens. Matter {\bf 18} L355 (2006)
\bibitem{HongAPL2006} N. Hoa Hong, N. Poirot and J. Sakai, Appl. Phys. Lett. {\bf89} 042503 (2006)
\bibitem{Schwartz2004} D. A. Schwartz, and D. R. Gamelin, Adv. Mater. {\bf16} 2115 (2004)
\bibitem{Radovanovic} P. V. Radovanovic and D. R. Gamelin, Phys. Rev. Lett. {\bf91} 157202 (2003)
\bibitem{HongPRB2005} N. Hoa Hong, J. Sakai, N. T. Huong, N. Poirot and A. Ruyter, Phys. Rev. B {\bf72} 45336 (2005)
\bibitem{CNR} A. Sundaresan, R. Bhargavi, N. Rangarajan, U. Siddesh and C.N.R. Rao, Phys. Rev. B. {\bf74} 161306R (2006)
\bibitem{HongJPC2007} N. H. Hong, J. Sakai and V. Brize, J. Phys: Condens. Matter {\bf19} 036219(2007)
\bibitem{Sreedharan} V. Sridharan, S. Banerjee, M. Sardar, S. Dhara, N. Gayathri and V. S. Sastry, cond - mat/0701232
\bibitem{condmatznomn} S. Banerjee, K. Rajendran, N. Gayathri, M. Sardar, S. Senthilkumar, V. Sengodan, arXiv:0704.3541v1,v2 
\bibitem{Garcia} M.A.Garcia, M.L.Ruiz-Gonzalez, A.Quesaga, J.L.Costa-Kramer, J.F.Fernandes, S.J. Khatib, A. Wennberg, A.C. Caballero, M.S.Martin-Gonzalez, M.Villegas, F.Bronies, J.M.Gonzalez-Calbert and A.Hernando, Phys. Rev. Letts., {\bf 94}, 217206 (2005) 
\bibitem{JAP97}J.P.Wang, H. M. Zhong, Z.F.Li and Wei Lu, Jl. Appl. Phys. {\bf 97}, 086105 (2005) 
\bibitem{Wang} J.B.Wang, G.J.Huang, X.L.Zhong, L.Z.Sun, Y.C.Zhou and E.H.Liu, Appl. Phys. Letts. {\bf88} 252502 (2006)
\bibitem{Sharma}P.Sharma, A.Gupta, F.J. Owens and A. Inoue, K.V.Rao, Jl. Magn. Mag. Mats.{\bf282} 115 (2004)
\bibitem{Chen} W.Chen, L.F.Zhao, Y.Q.Wang, J.H.Miao, S.Liu, Z.C.Xia and S.L.Yuan, Appl. Phys. Letts., {\bf87} 042507 (2005)
\bibitem{Hou} D-L.Hou, X-J.Ye, H-J.Meng, H-J.Zhou, X-L Li, C-M Zhen, G-D Tang, Mat. Sc. Engg. B {\bf138} 184 (2007)
\bibitem{Ram} G.Lawes, A.S.Risbud, A.P.Ramirez and Ram Seshadri, Phys. Rev. B. {\bf71} 045201 (2005)
\bibitem{Alaria} J.Alaria, P. Turek, M. Bernard, M. Bouloudenine, A. Berbadj, N. Brihi, G. Schmerber, S. Colis and A. Dinia, Chem. Phys. Letts., {\bf415} 337 (2005)
\bibitem{Fukumura} T. Fukumura, Z. Jin, M. Kawasaki, T. Shono, T. Hasegawa, S. Koshihara and  H. Koinuma, Appl. Phys. Letts. {\bf78} 958 (2001)
\bibitem{Kolesnik} S. Kolesnik and B. Dabrowski, Jl. Appl. Phys. {\bf96} 5379 (2004)
\bibitem{condmatchar} K.Rajendran, S.Banerjee, S.Senthilkumaar, T.K.Chini, V.Sengodan, cond-mat/0709.3901 (also submitted for publication)
\bibitem{Han2} S-J.Han, T-J.Jang, Y.B.Kim, B-G.Kim, B.G. Park, J. H. Park, Y. H. Jeong, Appl. Phys. Letts., {\bf83}, 920 (2003)
\bibitem{Chen2} W.Chen, L.F.Zhao,Y.Q.Wang, J.H.Miao, S. Liu, Z.C. Xia and S. L. Yuan, Solid. State. Coom. {\bf134} 827 (2005)
\bibitem{Spalek} J. Spaleck, A. Lewicki, Z. Tarnawski, J.K.Furdyana, R.R. Galazka and Z. Obuszko, Phys. Rev. B. {\bf33} 3407 (1986)
\bibitem{Luo} J. Luo, J.K. Liang, Q.L.Liu, F.S. Liu, Y.zhang, B.J. Sun and G. H. Rao, Jl. Appl. Phys. {\bf 96} 086106 (2005)
\bibitem{Dhar} S. Dhar, O. Brandt, M. Ramsteiner, V. F. Sapega and K. H. Ploog, Phys. Rev. Letts. {\bf94} 037205 (2005)
\bibitem{Orlov} A. F. Orlov, N. S. Perov, L. A. Balagurov, A. S. Konstantinova and D. G. Yarkin, JETP Letts. 352 (2007)
\bibitem{Ogale} S. B. Ogale, R. J. Choudhary, J. P. Buban, S. E. Lofland, S. R. Shinde, S. N. Kale, V. N. Kulkarni, J. Higgins, C. Lanci, J. R. Simpson, N. D. Browning, S. Das Sarma, H. D. Drew, R. L. Greene, and T. Venkatesan, Phys. Rev. Letts. {\bf91} 077205 (2003)
\bibitem{Song}C. Song, K. W. Geng, F. Zeng, X. B. Wang, Y. X. Shen, F. Pan,Y. N. Xie, T. Liu, H. T. Zhou, and Z. Fan, Phys. Rev. B {\bf73}, 024405 (2006)
\bibitem{Priya} Priya Mahadevan, J. M. Osorio-Guillen and Alex Zunger, Appl. Phys. Letts {\bf86} 172504 (2005)

\end{thebibliography}
\end{document}